\documentclass[aps,pra,reprint]{revtex4-1}
\usepackage[svgnames]{xcolor}
\usepackage{amsmath}
\usepackage{bm}
\usepackage{braket}
\usepackage{listings}
\usepackage{amssymb}
\usepackage{graphicx}
\usepackage{comment}
\graphicspath{{./img/}}

\begin{document}
\title{Hierarchical Schr\"{o}dinger Equations of Motion for Open Quantum Dynamics}
\author{Kiyoto Nakamura}
\email{nacamula@kuchem.kyoto-u.ac.jp}
\author{Yoshitaka Tanimura}
\email{tanimura.yoshitaka.5w@kyoto-u.jp}
\affiliation{Department of Chemistry, Graduate School of Science, Kyoto University, Sakyoku, Kyoto 606-8502, Japan}
\date{\today}

\begin{abstract}
We rigorously investigate the quantum non-Markovian dissipative dynamics of a system coupled to a harmonic-oscillator bath by deriving hierarchical Schr\"{o}dinger equations of motion (HSEOM) and studying their dynamics. The HSEOM are the equations for wave functions derived on the basis of the Feynman-Vernon influence functional formalism for the density operator, $\braket{q|\rho(t)|q'}$, where $\bra{q}$ and $\ket{q'}$ are the left- and right-hand elements.  The time evolution of $\bra{q}$ is computed from time $0$ to $t$, and, subsequently, the time evolution of $\ket{q'}$ is computed from time $t$ to $0$ along a contour in the complex time plane.
By appropriately choosing functions for the bath correlation function and the spectral density, we can take advantage of an HSEOM method to carry out simulations without the need for a great amount of computational memory.
As a demonstration, quantum annealing simulation
for a ferromagnetic $p$-spin model is studied.
\end{abstract}
\maketitle

\section{INTRODUCTION}
Open quantum systems have been a subject of interest for many years\cite{LeggettRMP87,Grabert88}.
A great deal of effort has been dedicated to numerically calculating the time evolution of such model systems, and
several numerically rigorous approaches have been developed for studying spin-boson systems and Brownian oscillator systems.
These approaches include the quasiadiabatic propagator path integral (QUAPI)\cite{Makri2014}, the density matrix renormalization group (DMRG)\cite{AlexPlenio}, and the
reduced hierarchical equations of motion (HEOM) methods\cite{heomHigh,TanimuraPRA90,heom,TanimuraJPSJ06,TJCP2012,Tanimura2014,Tanimura2015,Shi09,YanPade10A, TJCP2012,Cao13sto,Wu15eHeom,Kleine16chev,DueanCao17,Yan2012,Shi14, Chen15}.
Although the relaxation processes exhibited by a model system under external perturbations are now well understood, these processes of complex systems consisting of many energy states and/or potential-energy surfaces defined in multidimensional configuration spaces have not been thoroughly explored due to a lack of computational power. This is due to the fact that the quantum dynamics of an open $N$-state system must be described using an $N \times N$ reduced density matrix in order to have time-irreversible processes described by a non-Hermitian propagator, while the quantum dynamics of an isolated $N$-state system are described using an $N \times 1$ column or a $1 \times N$ row vector. Moreover, if we consider systems that are strongly influenced by heat baths and
need to adopt a nonperturbative approach, such as HEOM\cite{ SakuraiJPC11,KatoJPCB13,SakuraiJPSJ13,IkedaJCP17,kramerGPU,kramer,Tsuchimoto,CaoGHEOM1,CaoGHEOM2},
more computational resources are necessary.
For this reason, the memory required to compute the density matrix elements becomes a serious issue when studying large systems.

\textcolor{black}{Methodologies based on wave functions for the full Hamiltonian have been developed in order to avoid the reduced description of the system.
The multiconfiguration time-dependent Hartree (MCTDH) approach\cite{ML-MCTDH1,ML-MCTDH2,WangTHoss1,WangTHoss2} employs time-dependent basis sets to represent the total wave function. Then, a variational principle is applied to derive the optimal equation of motion in order to reduce the bath degrees of freedom.
This kind of approach has wider applicability than the reduced equation of motion. With the MCTDH approach, we can treat nonlinear system-bath couplings and anharmonic bath modes\cite{WangTHoss2}, which cannot be treated with the conventional HEOM approach. However, the number of bath modes must be increased until convergence is reached. This implies that the study of long-time behavior requires more basis sets, which makes the calculation more difficult.
Moreover, the time evolution obtained with the wave-function-based approach describes time-reversible processes and, thus, the thermal equilibrium state cannot be obtained with this approach.}

The stochastic unraveling method\cite{Strunz14sto, Strunz15sto,Strunz17sto,Zhao16sto,Zhao17sto} is another approach to reduce the computational costs of simulations for open quantum systems. While the stochastic trajectories obtained from this approach are useful to analyze a role of noise, the efficiency of calculations is not always high because the sampling of stochastic variables is not simple, and many of these approaches have to employ auxiliary variables, including auxiliary density operators (ADOs), in addition to stochastic variables.

In the present paper, we derive hierarchical Schr\"{o}dinger equations of motion (HSEOM) for wave functions.
The time evolution of the reduced density matrix elements can be obtained by numerically integrating the HSEOM with respect to $t$ along the contour in the complex time plane with the aid of a bath correlation function expressed in terms of a set of special functions. This expression maintains the stability of the HSEOM while integrating along the contour in the direction of decreasing time.

\section{Hierarchical Schr\"{o}dinger Equations of Motion}
We consider a system $S$ coupled to a bath $R$ of harmonic oscillators. The Hamiltonian of the total system is given by\cite{LeggettRMP87,Grabert88}
\begin{align}
\hat H_{tot} &= \hat H_S - \hat V\sum\limits_j {k_j \hat x_j} + \sum\limits_{j} {\left( \frac{\hat p_j^2 }{2m_j } + \frac{1}{2} m_j \omega _j^2 \hat x_j^2 \right) },
\end{align}
where $\hat H_S$ is the Hamiltonian of the system and $\hat V$ is the system part of the system-bath interaction.
The bath degrees of freedom are treated as an ensemble of harmonic oscillators,
with the momentum, position, mass, and frequency of the $j$th bath oscillator given by $\hat{p}_{j}$, $\hat{x}_{j}$, $m_{j}$, and $\omega_{j}$, respectively. The quantity $k_j$ is the coupling constant for the interaction between the system and the $j$th oscillator.
The heat bath is characterized by the spectral density function, defined as
\begin{align}
J(\omega) \equiv \sum_{j }\left( \frac{k_{j}^2}{2m_{j}\omega_{j}}\right) \delta(\omega-\omega_{j}),
\end{align}
and the inverse temperature $\beta \equiv 1/k_{\mathrm{B}}T$, where $k_\mathrm{B}$ is the Boltzmann constant.

\begin{figure}[t]
\centering
\includegraphics[scale=0.7]{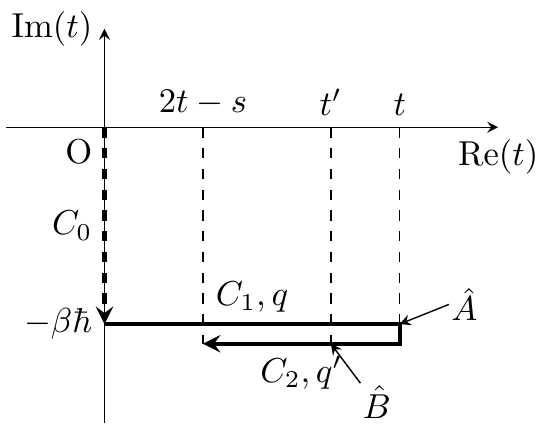}
\caption{The contour $C$ in the complex time plane. \label{fig:contour2}}
\end{figure}
We consider a multitime correlation function,
\begin{align}
\Psi_{AB}(t; t')= \mathrm{tr}\{ \hat{A}(t) \hat{\rho}_{tot}(0) \hat{B}(t') \},
\end{align}
where $\hat A$ and $\hat B$ are operators acting on the system $S$, and $\hat{\rho}_{tot}(0)$ is the initial state of the  density operator for the total system.
For the contour illustrated in Fig. \ref{fig:contour2}, we can express the correlation function in path-integral form. In order to simplify the derivation, we adopt factorized initial conditions at $t=0$ as $\hat {\rho}_{tot}(0)=\hat{\rho}_{S}(0) \otimes \hat{\rho}^{eq}_{R}$ and do not consider the time evolution from $t=0$ to $0-i \beta \hbar$ along the contour $C_0$\cite{Tanimura2014,Tanimura2015}. If necessary, the thermal equilibrium state for the total system can be obtained from these conditions by integrating the HSEOM for a sufficiently large time $t$.
Because the operators depend only on the system variable as ${A(\hat{q})}$, ${B(\hat{q})}$, and $V(\hat{q})$, where $\hat{q}$ is the position  or spin operator, we can trace over the bath degrees of freedom.
For a harmonic heat bath, we can evaluate $\Psi_{AB}(t;t')$ analytically as\cite{ TanimuraJPSJ06,Tanimura2014,Tanimura2015}
\begin{align}
\Psi_{AB}(t; t')=& \int dq_{i}' \int dq_{i} \int _{C} \mathcal{D}[\tilde{q}(\cdot)] A(q, t)B(q', t') \nonumber \\
&\times \exp \left[ \frac{i}{\hbar} \int_{C} d\tau L_{S}(\dot{\tilde{q}}, \tilde{q}, \tau) \right. \nonumber \\
&- \frac{1}{{\hbar}^2} \left. \int _{C} d\tau \int_{C'} \hspace{-1pt} d\tau' V(\tilde{q}, \tau) \alpha(\tau-\tau')
V(\tilde{q}, \tau') \right] \nonumber \\
&\times
 \braket{ q_{i} | \hat{\rho}_{S}(0) | q_{i}'}.
\end{align}
where $L_{S}(\dot{\tilde{q}}, \tilde{q}, \tau)$ is the Lagrangian for the system Hamiltonian $\hat H_S$ and the path integral is indicated by $\int \mathcal{D}[\tilde{q}(\cdot)]$. The contour $C'$, along which the integration over  $\tau'$ is carried out, is $C$ up to $\tau$. The integrals here are carried out in the direction of the arrow in Fig. \ref{fig:contour2}. The position of the system along the contour $\tilde{q}$ is $q$ or $q'$, depending on whether the contour integral is along $C_{1}$ or $C_{2}$.
The variables $A(q, t)$, $B(q', t') $, and $V(\tilde{q}, \tau)$ are the path-integral representation of ${A(\hat{q})}$, ${B(\hat{q})}$, and $V(\hat{q})$ respectively. For the case of a half-spin system, see the Appendix.
The bath correlation function $\alpha(t)$ is defined as
\begin{align}
\label{eq:bathcorr}
\alpha(t) \equiv \hbar \int_{0}^{\infty} d \omega J(\omega) \left\{
\coth\left(\frac{\beta \hbar \omega}{2}\right) \cos \omega t - i \sin \omega t \right\}
\end{align}
We then rewrite the correlation function as $\Psi_{AB}(t; t') = \int dq_{i}' \braket{ q_{i}'|  \phi_{C} (q'_{i}) }$, where $  \ket{\phi_{C} (q'_{i})} $ is the reduced wave function (RWF) integrated along the contour with the initial wave function $\hat{\rho}_{S}(0)\ket{q'_{i}}$. We can also evaluate the correlation function in terms of an energy-state representation as $\Psi_{AB}(t;t') = \sum_{n'_{i}} \braket{n'_{i}|\phi_{C}(n'_{i})}$. In the following, we derive the equations of motion for $\ket{ \phi_{C} (n'_{i})}$ in hierarchical form.

The HEOM have been derived for Drude-type\cite{heomHigh, TanimuraPRA90,heom,TanimuraJPSJ06,Tanimura2014,Tanimura2015,
Shi09,YanPade10A,Yan2012}, Brownian-type\cite{TJCP2012}, and Lorentzian-type\cite{kramer} spectral densities by expressing $\alpha(t)$ as a sum of exponential functions as $\alpha(t)=\sum_k c_k e^{-a_k|t|}$, where $c_k$ and $a_k$ are real or complex constants. Here, we express the bath correlation function using special functions $\{\varphi_{k}(t) \}$\cite{Wu15eHeom, DueanCao17,Kleine16chev} because the exponential form of $\alpha(t)$ becomes unstable in the time integration along the returning contour $C_2$.
The bath correlation function is now expressed as
$\alpha(t) = \sum_{k = 0} ^{K-1} c_{k} \varphi_{k} (t)$, where $c_{k}$ are complex constants and
the number of basis elements is restricted to some value $K$ to facilitate the numerical computations.
In order to obtain a closed set of equations, we choose the set of special functions $\{ \varphi_{k}(t) \}$ so as to ``approximately'' satisfy the relation
\begin{align}
\frac{d}{dt} \varphi_{k}(t) = \sum_{k'=0}^{K-1} \eta_{k, k'} \varphi_{k'} (t),
\end{align}
where $\eta_{k,k'}$ are the expansion coefficients. Although this type of decomposition requires more hierarchical terms than the conventional HEOM formalism employing the exponential-function decomposition scheme, in particular to study long-time behavior, this allows us to study wider classes of spectral densities at any temperature, including a sub-Ohmic spectral density at zero temperature\cite{DueanCao17}.
Using the above relation, we obtain the HSEOM by differentiating the RWF and auxiliary wave functions (AWFs) along the contours $C_{1}$ and $C_{2}$ in the same way as in the conventional HEOM approach\cite{TanimuraJPSJ06, Tanimura2014,Tanimura2015}.
They take the following form:
\begin{align}
\frac{\partial}{\partial s}  \ket{\phi_{\vec{n}} (s; n'_{i})} = &
\mp \frac{i}{\hbar} \hat{H}_{S} \ket{\phi_{\vec{n}} (s; n'_{i}) } \nonumber \\
&
\pm \sum_{k=0}^{K-1} \sum_{k'=0}^{K-1} \eta_{k, k'} n_{k}
\ket{ \phi_{\vec{n} - \vec{e}_{k} + \vec{e}_{k'}}  (s; n'_{i})} \nonumber \\
& \mp \frac{i}{\hbar} \hat{V} \sum_{k=0}^{K-1} c_{k} \ket{  \phi_{\vec{n} + \vec{e}_{k}}  (s; n'_{i})} \nonumber \\
&
\mp \frac{i}{\hbar} \hat{V} \sum_{k=0}^{K-1} n_{k} \varphi_{k} (0)
\ket{  \phi_{\vec{n} - \vec{e}_{k}}  (s; n'_{i})},
\label{HSEOMC1}
\end{align}
and the AWFs are expressed in terms of a line integral as follows:
\begin{gather}
 \ket{ \phi_{\vec{n}} (s; n'_{i}) } =
 \sum_{n, n_{i}} \ket{n} \int dq \braket{n|q}   \int dq_{i} \int _{0}^{s} \mathcal{D} [\tilde{q}(\tau(\cdot))] \nonumber \\
\begin{aligned}[b]
&\times
\prod_{k=0}^{K-1}
\left(\hspace{-1pt}-\frac{i}{\hbar} \int _{0}^{s} \hspace{-7.5pt}ds'' \frac{d\tau(s'')}{d s''}
\varphi_{k}(\tau(s)-\tau(s'')) V(\tilde{q}, \tau(s'')) \hspace{-1pt} \right) ^{n_{k}} \\
&\times \exp\left[ \frac{i}{\hbar} \int _{0}^{s} ds' \frac{d\tau(s')}{ds'}
L_{S}(\dot{\tilde{q}}, \tilde{q}, \tau(s'))  \right] \mathcal{F}(s, V) \\
&\times
\braket{q_{i}|n_{i} } \braket{ n_{i} |\hat{\rho}_{S}(0)| n'_{i} },
\end{aligned}
\end{gather}
\textcolor{black}{where the influence functional is expressed as
\begin{gather}
\mathcal{F} (s, V) = \exp \left[-\frac{1}{{\hbar}^2} \int_{0}^{s} ds' \frac{d\tau(s')}{ds'}
\int_{0}^{s'} ds'' \frac{d\tau(s'')}{ds''} \right. \nonumber \\
\left. \times V(\tilde{q}, \tau(s')) \sum_{k=0}^{K-1} c_{k} \varphi_{k}(\tau(s')-\tau(s'')) V(\tilde{q}, \tau(s''))  \right],
\end{gather}}
and we have introduced the time variable $\tau(s)$ for $0 \leq s \leq 2t$ defined as
\begin{align}
\tau(s) \equiv \left\{
\begin{array}{lll}
  s, & 0 \leq s \leq t, & \mbox{for $C_{1}$}\\
  2t-s, & t \leq s \leq 2t, & \mbox{for $C_{2}$}.
\end{array}
\right .
\end{align}
Accordingly, we define $\tilde{q}$ such that $\tilde{q} = q$ for $0 \leq s \leq t$ and
$\tilde{q} = q'$ for $t \leq s \leq 2t$.

In Eq. \eqref{HSEOMC1}, the upper signs of $\mp$ and $\pm$ correspond to $0 \leq s \leq t$, while the lower signs correspond to $t \leq s \leq 2t$.
The vector $\vec{n} = (n_{0}, n_{1}, \ldots , n_{k}, \ldots, n_{K-1})$, used as a subscript here, distinguishes the AWFs and $\ket{\phi_{\vec{n} = \vec{0}}(s;n'_{i})}$ corresponds to the RWF. Each $n_{k}$ is a non-negative integer, and $\vec{e}_{k}$ is the unit vector of the $k$th element. The level of the hierarchy, $N$, is given by $N = \sum_{k=0}^{K-1} n_{k}$. We choose a maximum value of this level, $N_{\max}$. Any AWF whose level is higher than $N_{\max}$ is set to zero, in order to obtain a closed set of equations.

In order to compute the correlation function, we integrate Eq. \eqref{HSEOMC1} with the upper signs from the initial wave function, $\ket{ \phi _{\vec{n}}  (s=0; n'_{i})} $, up to the time $s=t$. At $t$, we apply the operator $\hat{A}$ to $\ket{ \phi_{\vec{n}} (s=t; n'_{i})} $ as $\hat{A} \ket{ \phi_{\vec{n}} (s=t; n'_{i}) } \rightarrow \ket{\phi_{\vec{n}} (s=t; n'_{i}) } $.
Then, after the time integration of Eq. \eqref{HSEOMC1} with the lower signs along the contour $C_2$ up to the time $s = 2t - t'$, we apply the operator $\hat{B}$ as $\hat{B} \ket{ \phi_{\vec{n}} (s=2t-t'; n'_{i})} \rightarrow \ket{ \phi_{\vec{n}} (s=2t-t'; n'_{i})}$. We obtain a wave function with a fixed initial wave function $\ket{\phi_{\vec{0}}(s=2t; n'_{i})}$ after continuing the integration up to $s=2t$. For the calculations of the two-body correlation function, we must iterate the above-mentioned calculation with all the different initial states $\ket{n'_{i}}$. A possible number of initial states is equivalent to the number of system states, and hence we must iterate the calculation $N$ times for an $N$-state system. We then obtain the two-body correlation function with the equation
\begin{align}
\Psi_{AB}(t;t') = \sum_{n'_{i}} \braket{ n'_{i}|  \phi_{\vec{0}}(s=2t; n'_{i})}.
\end{align}
 For the calculations of the equilibrium correlation function, we can start from any initial state, with the time $t'$, at which $\hat{B}$ is applied to the system, chosen to be sufficiently large, because the steady-state solution of the HSEOM is a correlated equilibrium state\cite{Tanimura2014,Tanimura2015}.
The number of operators to be applied is not restricted to two, and we can evaluate the higher-order nonlinear response functions by increasing this number\cite{TanimuraJPSJ06}.
We can also evaluate the reduced density matrix elements, $\braket{i| \mathrm{tr}_{R}\{\hat{\rho}_{tot}(t)\} |j} $, with $\hat B$ chosen to be the unit operator and $\hat{A}$ chosen such that $\hat{A} = \ket{j}\bra{i}$.

Here, in order to reduce the number of iterations, we introduce a ``localized initial state'' of the form $\hat{\rho}_{S}^{loc}(0) = \ket{k}\bra{k}$. This initial state enables us to evaluate the two-body correlation functions with the equation
\begin{align}
\Psi_{AB}(t;t') = \braket{k|\phi_{\vec{0}}(s=2t; n'_{i} = k)},
\end{align}
reducing the number of iterations $N$ to $1$ for $N$-state systems.
By appropriately choosing the transformation matrix $\hat{C}$, we can calculate the two-body correlation function from the desired initial conditions by using the result from the localized initial state $\hat{\rho}_{S}^{loc}(0)$ with the equation
$\Psi_{AB}(t;t') = \mathrm{tr}\{\hat{A}(t) \hat{C}(0) \hat{\rho}_{S}^{loc} \hat{C}^{\dagger}(0)\hat{B}(t')\}$.

\section{numerical results}
We now report the results of numerical simulations that demonstrate the applicability and the validity of the HSEOM, given in Eq. \eqref{HSEOMC1}. We first consider the spin-boson case, with the system Hamiltonian given by
\begin{align}
\label{spinBoson}
\hat{H}_{S} = -\frac{1}{2}\hbar \omega_{0}\hat{\sigma}^{z},
\end{align}
and the system part of the system-bath interaction given by
\begin{align}
\hat{V} = -\frac{1}{2}\hbar \hat{\sigma}^{x},
\end{align}
where $\hat{\sigma}^{x}$ and $\hat{\sigma}^{z}$ are Pauli matrices.

We tested Bessel functions\cite{Kleine16chev} and harmonic-oscillator wave functions\cite{Wu15eHeom} to express the bath correlation function $\alpha(t)$. We found that Bessel functions allow for a more efficient treatment than the harmonic-oscillator wave functions for an Ohmic spectral density. In terms of Bessel functions of the first kind, $J_{k}(t)$, the bath correlation function is expressed as
\begin{align}
\alpha(t) =  \sum_{k=0}^{K-1} c_{k} J_{k}(\Omega t),
\label{specialfunc}
\end{align}
and $c_k$ are approximated as
\begin{align}
\label{coefficient}
c_{k}\equiv \hbar \Omega
\int_{-1}^{1} dx (2 - \delta_{0, k})(-i)^{k} T_{k}(x)\frac{J(\Omega x)}{1- e^{-\beta \hbar \Omega x}}.
\end{align}
\textcolor{black}{
This equation is derived with the aid of the Jacobi-Anger identity\cite{Guanhua12chev, Kleine16chev},
\begin{gather*}
e^{-i \Omega x t} = J_{0}(\Omega t) + \sum_{k=1}^{\infty} 2(-i)^{k} T_{k}(x) J_{k}(\Omega t), \nonumber \\
\forall t \in \mathbb{R}, \forall x \in [-1, 1].
\end{gather*}
Here, $T_{k}$ is a Chebyshev polynomial and in order to use this identity we have modified Eq. \eqref{eq:bathcorr} in the following form:
\begin{align}
\alpha(t) 
& = \hbar \int_{-1}^{1} dx \frac{J(\Omega x) e^{-i \Omega x t}}{1 - e^{- \beta \hbar \Omega x}} .
\label{eq:bathcorrmod}
\end{align}
We can reduce the range of the integration in Eq. (\ref{eq:bathcorrmod}) from $(-\infty, \infty)$ to $[-\Omega, \Omega]$ with the appropriate cutoff frequency $\Omega$ because the spectral density decays to $0$ as $|\omega| \to \infty$.
In the zero-temperature limit ($\beta \hbar \to \infty$), Eq. \eqref{eq:bathcorrmod} is not exact,
and for this case we evaluate $c_{k}$ using the following equations instead of Eq. \eqref{coefficient}:
for the real part (even $k$),
\begin{align}
c_{k} = \frac{\hbar \Omega}{2} \int_{-1}^{1} dx (2 - \delta_{0, k})(-i)^{k} T_{k}(x) \mathrm{sgn}(x) J(\Omega x),
\end{align}
and for the imaginary part (odd $k$),
\begin{align}
c_{k} = \frac{\hbar \Omega}{2} \int_{-1}^{1} dx (2 - \delta_{0, k})(-i)^{k} T_{k}(x) J(\Omega x).
\end{align}}
We can differentiate $J_{k}(t)$ with respect to $t$ by setting
\begin{align}
\eta_{k, k'} = & \left \{
\begin{array}{ll}
\Omega/2 & (k' = k -1)\\
-\Omega/2 & (k' = k + 1),
\end{array}
\right .
\end{align}
$\eta_{0, 1} = -\Omega $, and, otherwise, $\eta_{k, k'} = 0$.  Here, we have approximated the time derivative of the $(K-1)$th function as
$d J_{K-1}(\Omega t) /dt  \simeq \Omega J_{K-2}(\Omega t) / 2$
for the sake of numerical computations.
\textcolor{black}{This approximation is accurate when we choose the number of Bessel functions $K$ to satisfy the condition $J_{K}(\Omega t) \simeq 0$ for all $t \in [0, T]$, where $T$ is the length of the simulation time.
This indicates that even when the bath correlation function is expressed with a small number of Bessel functions, we need to set a larger value for $K$ for longer time simulations.
By contrast, in the case of the conventional HEOM,  the number of exponential functions does not depend on $T$.}

We considered the Ohmic spectral density with the exponential cutoff
\begin{align}
J(\omega) =\eta \omega e^{-|\omega| / \gamma},
\end{align}
and the circular cutoff \cite{Ando98, Guanhua12chev}
\begin{align}
J(\omega) = \zeta \omega\sqrt{1 - (\omega/\nu)^{2}},
\end{align}
where $\gamma$ and $\nu$ are the cutoff frequencies. We find that if we choose $\nu = \gamma$ and $\zeta = 2 \eta / e$, where $e$ is the base of the natural logarithm, the numerical results obtained with these two types of cutoffs exhibit similar behavior under the condition $\omega_{0} \ll \gamma$, where $\omega_{0}$ is the characteristic frequency of the system with the Hamiltonian given in Eq. \eqref{spinBoson}. For the exponential cutoff, we set $\Omega$ manually with the condition $\gamma < \Omega$,  and for the circular cutoff,  we evaluated $\Omega$ analytically as $\Omega = \nu$.

It should be noted that in the case of the Ohmic spectral density with the circular cutoff, the imaginary part of the bath correlation function defined by $\alpha(t)=\alpha'(t)-i\alpha''(t)$ is analytically evaluated as\cite{Ando98}
\begin{align}
\alpha''(t) =  c_1 J_{1}(\nu t) + c_3 J_{3}(\nu t),
\end{align}
where $c_1=c_3=\pi \hbar \zeta \nu^{2}/{8}$. In the high-temperature limit $\beta \hbar \to 0$, the real part of the bath correlation function reduces to $\alpha'(t) = {\pi} {\zeta \nu}(J_{0}(\nu t) + J_{2}(\nu t))/2\beta$.
This indicates that for the construction of the HSEOM, the circular cutoff allows for a more efficient approach than the exponential cutoff.

\begin{figure}[t]
\centering
\includegraphics[scale=0.5]{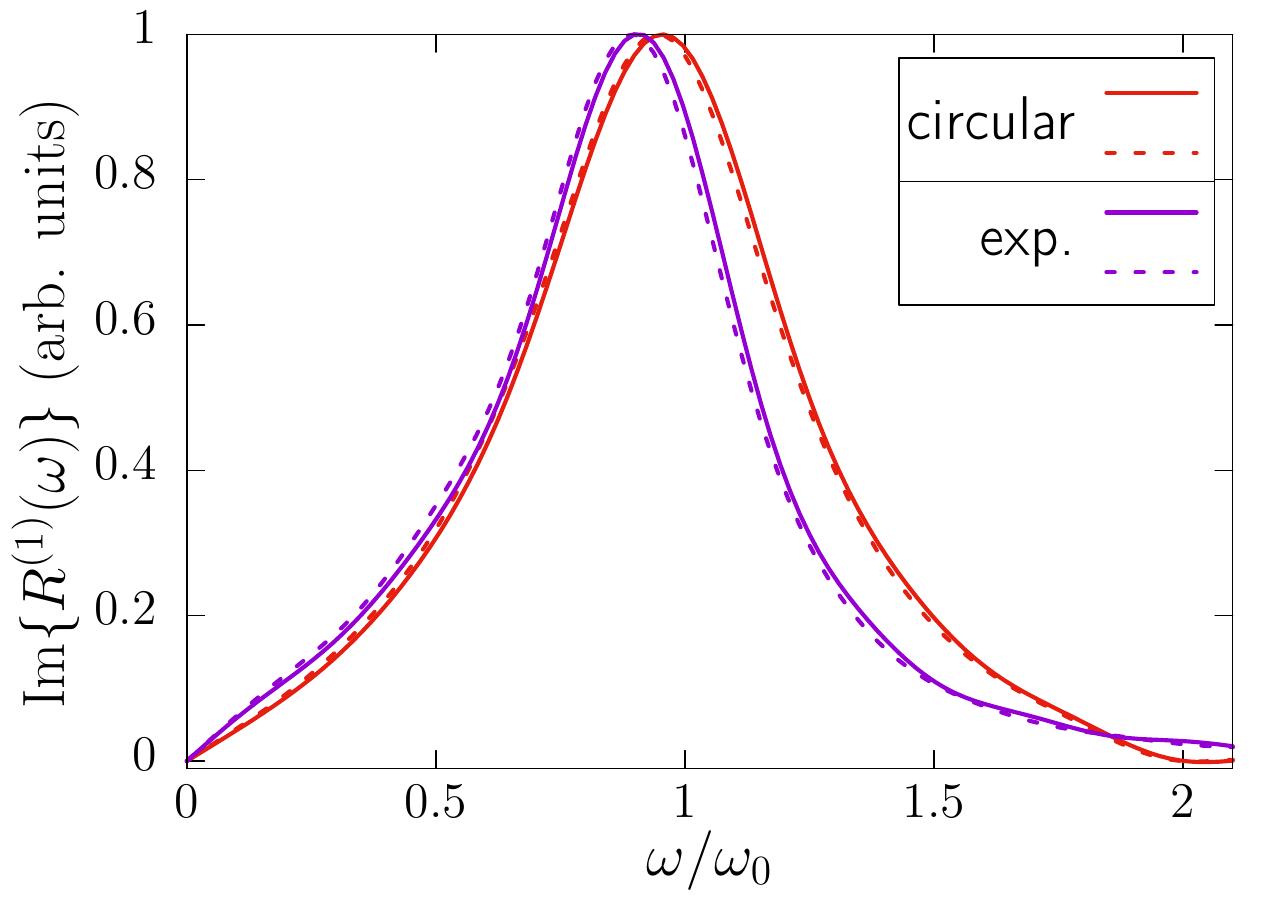}
\caption{The imaginary part of the Fourier transform of the first-order response function $R^{(1)}(\omega)$ for a spin-boson system with an Ohmic spectral density with a circular cutoff (red) and an exponential (exp.) cutoff (blue).
Results for two cases, in which $\beta \hbar = 3$ (solid curve) and $\beta \hbar \to \infty$ (zero-temperature case, dashed curve) are displayed.
The unit of the frequency $\omega$ is set to the characteristic frequency of the system $\omega_{0}$ in Eq. \eqref{spinBoson}.
\label{fig:response}}
\end{figure}

In Fig. \ref{fig:response}, we depict the imaginary part of the Fourier transform of the first-order response function for the spin-boson system,
$R^{(1)}(\omega) = \int_{0}^{\infty} dt e^{-i \omega t}R^{(1)}(t)$, where $R^{(1)}(t)$ is defined by
\begin{align}
\label{eq:response}
R^{(1)}(t) &\equiv -\frac{i}{2}\braket{[\hat{\sigma}^{x}(t), \hat{\sigma}^{x}]}_{eq} \nonumber \\
&= \mathrm{Im} [\Psi_{\sigma^{x} \sigma^{x}}(t;t_{0})].
\end{align}
We can employ the localized initial state as $\hat{\rho}_{S}^{loc}(0) = \ket{1}\bra{1}$ and reduce the number of iterations
because the total equilibrium state is obtained at time $t_{0}$ from this initial state.
In the numerical calculations, we chose the system parameters as $\omega_{0}=\pi$, and the bath parameters as $\hbar \zeta= 0.35, \nu=6$, and $\Omega = 20$ in the case of the exponential cutoff and $\Omega = 6$ in the case of the circular cutoff. Two temperature cases  ($\beta \hbar =3$ and $\beta \hbar \to \infty$) are depicted in Fig. \ref{fig:response}. In both cases,
the number of basis elements, $K$, was set to $80$ for the exponential cutoff and $20$ for the circular cutoff, and the maximum level of the hierarchy, $N_{\max}$, was set to $3$. The number of AWFs used to solve the HSEOM in the case of the exponential and circular cutoffs was $91881$ and $1771$, respectively. This indicates that the circular cutoff allows for a more efficient construction of the hierarchy, while the results are similar, as illustrated in Fig. \ref{fig:response}.

To confirm the numerical accuracy of our computations, we calculated the same variable using the extended hierarchical equations of motion (eHEOM)\cite{Wu15eHeom} approach with the same set of Bessel functions. The results obtained from the HSEOM and eHEOM are almost identical (not shown).
\textcolor{black}{
It should be noted that in the case that we use the same function set for approximating the bath correlation function, the eHEOM require more ADOs than the HSEOM because, with the eHEOM, we need two sets of special functions, one representing the real part and one representing the imaginary part of the bath correlation function, while we need only one set to represent Eq. \eqref{specialfunc} in the HSEOM case.
}
By contrast, the HSEOM require more time integrations than the eHEOM because we have to integrate along $C_1$ and $C_2$ consecutively in the HSEOM case, while these integrations can be carried out simultaneously in the eHEOM case.
\textcolor{black}{
Moreover, when we calculate over longer time periods $T$, both the HSEOM and the eHEOM require more AWFs or ADOs in order to obtain convergent results because the long-time behavior can be described only by a larger set of Bessel functions.
}

\begin{figure}[t]
\centering
\includegraphics[scale=0.5]{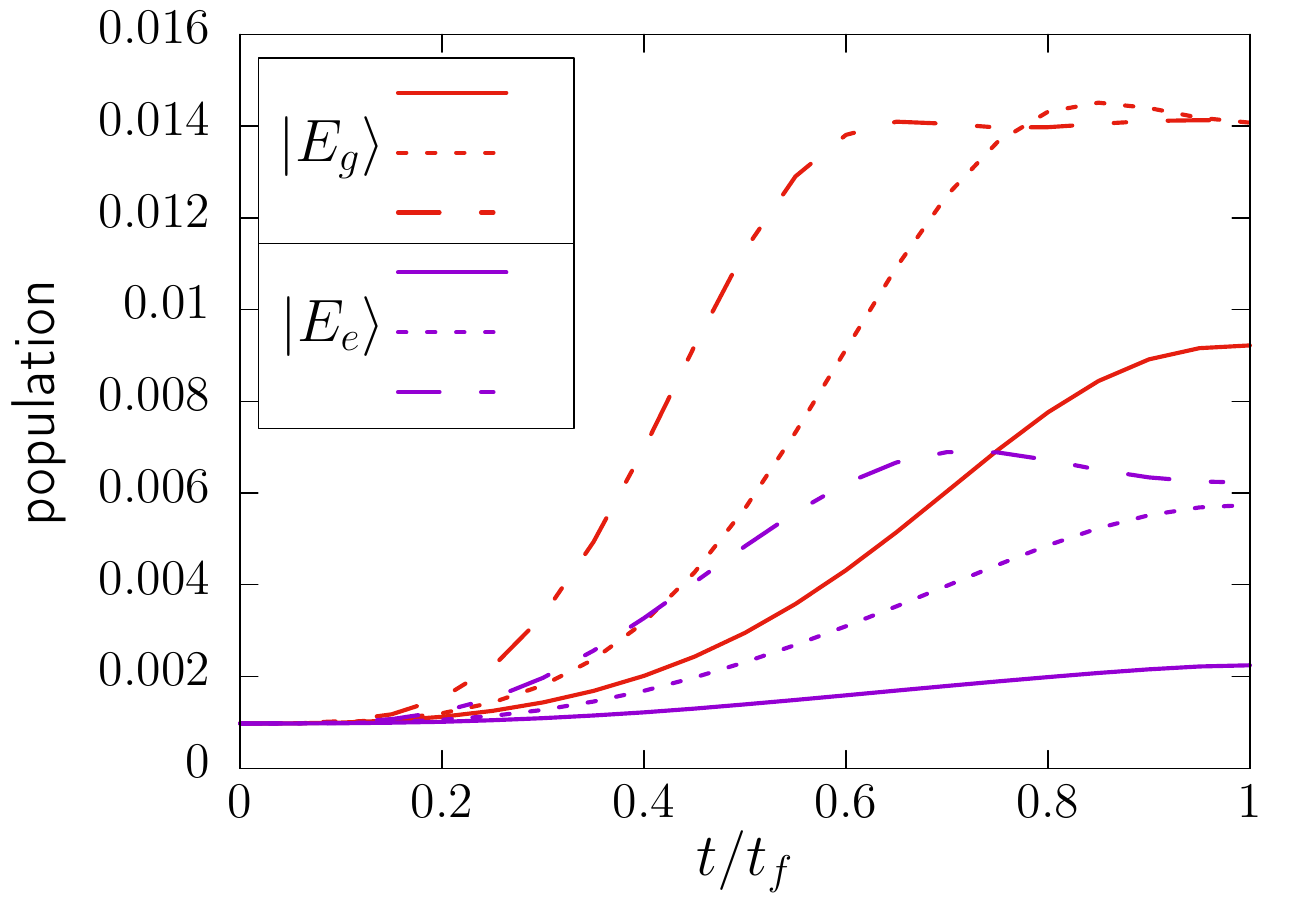}
\caption{The time evolution of the ground-state (red) and first-excited-state (blue) populations in the (i) weak $\hbar \zeta = 0.01$ (solid line), (ii) intermediate $\hbar \zeta = 0.1$ (dashed line), and (iii) strong $\hbar \zeta = 0.5$ (dot-dashed line) coupling cases calculated from the quantum annealing simulation at the zero temperature.
The unit of the time $t$ is set to $t_{f}$ in Eq. \eqref{eq:qa}.
\label{fig:qa}}
\end{figure}
Next we demonstrate quantum annealing\cite{Nishimori98} of a 10 qubit system for the ferromagnetic $p$-spin model. The system Hamiltonian is expressed as\cite{Chakra13QA}
\begin{align}
\hat{H}_{S}(t) = \left(1 - \frac{t}{t_f}\right) \hat{H}_{0} + \frac{t}{t_f}\hat{H}_{1},
\label{eq:qa}
\end{align}
where
\begin{align}
\hat{H}_{0} = -\hbar \Gamma \sum_{i=0}^{\mathcal{N}-1} \hat{\sigma}_{i}^{x},
\end{align}
and  $\Gamma$ is the magnitude of the transverse magnetic field in the $x$ direction and $\hat{\sigma}_{i}^{\alpha}$ ($\alpha=x, y,$ and $z$) is the Pauli operator of the $i$th site.  The targeting Hamiltonian where we want to find the ground state is given by\cite{Jorg10pspin}
\begin{align}
\hat{H}_{1} = - \hbar \mathcal{N} \left(\sum_{i=0}^{\mathcal{N}-1} \frac{\hat{\sigma}_{i}^{z}}{\mathcal{N}}\right)^{p}.
\label{eq:pspin}
\end{align}
The system part of the interaction is expressed as
\begin{align}
\hat{V} = \hbar \sum_{i=0}^{\mathcal{N}-1} \hat{\sigma}_{i}^{z}.
\end{align}
We chose the system parameters as $\mathcal{N} = 10, \Gamma = 1, p = 5$, and $t_{f} = 1$ and the bath parameters for the Ohmic spectral density with a circular cutoff as $\beta \hbar \to \infty$ (zero-temperature case), $\nu = 3$. We consider the (i) weak-, (ii) intermediate-, and (iii) strong-coupling cases, $\hbar \zeta=0.01$, 0.1, and 0.5, respectively.
\textcolor{black}{We carried out numerical calculation using a PC with 3.00 GHz Dual Intel Xeon CPU (total $24$ cores). The number of basis elements and the maximum level of the hierarchy were set to $K = 5, N_{\max} = 3$ for the (i) weak- and (ii) intermediate-coupling cases and $K = 5, N_{\max} = 5$ for the (iii) strong-coupling case. As mentioned before, we have to employ the larger number of Bessel functions $K$ for larger $T$. Here, we found that $K$ is proportional to $T$. In this simulation, the length of the simulation time $T$ was set to $1$ for all three cases, and thus we set the same value $K = 5$, while we had to set larger $N_{\max}$ for the (iii) strong-coupling case. As a result, the total numbers of AWFs used in the calculations were $N_{tot} = 56$ for (i) and (ii), and $252$ for (iii), which required $14.0$ and $26.3$ MB of computational memory, respectively. The computation time on the Xeon PC for (i) and (ii) was $3$ minutes, whereas that for (iii) was $14$ minutes. We note that $N_{\max}$ also depends on the characteristic frequency of the system, $\omega_c$.
Here, for the ferromagnetic $p$-spin model defined as Eq. \eqref{eq:pspin}, we have $\omega_c \propto \mathcal{N}$ because the energies of the ground state and the most unstable state are $-\hbar \mathcal{N}$ and $\hbar \mathcal{N}$, respectively, while the other energy states are almost $0$. Thus, in the case of the $4$ qubit system, we had less AWFs as $N_{\max}=2$ for (i) and (ii), and $N_{\max}=4$ for (iii), while $K$ did not change. }

As the initial conditions, we set all elements of the reduced density matrix to be the same, and hence the transformation matrix of the localized initial state is defined as $\braket{i|\hat{C}|j} = \delta_{j, 0}/32$.

In Fig. \ref{fig:qa}, we depicted the time evolution of the populations of the ground state, $\ket{E_g} = \bigotimes _{i=0}^{\mathcal{N}-1} \ket{1}_{i}$, and the first-excited-state,  $\ket{E_e} = \ket{0}_{j} \bigotimes_{i=0, i\neq j}^{\mathcal{N}-1} \ket{1}_{i}$ (for the definitions of $\ket{0}$ and $\ket{1}$, see the Appendix). As was reported in the previous study with a perturbative Markovian approach\cite{Luci18QA}, the ground-state population is larger for the intermediate system-bath coupling case than the weak-coupling case. We found, however, that the ground-state population already reaches the maximum values in the intermediate-coupling case, while the increase of the population is faster for the stronger-coupling case.
Although the demonstrated calculations are too small to be practical results, this is the largest annealing simulation in terms of a quantum dissipative approach. In addition, this simulation was carried out at zero temperature in a quantum mechanically rigorous manner merely using a personal computer. This is because, with the HSEOM, we can reduce the size of the density matrices from $N^2$ to $N$ for $N$-state systems.

\section{Concluding remarks}
We developed the HSEOM for wave functions utilizing contour integration.
\textcolor{black}{With this formalism, we can reduce the size of the density matrices from $N^2$ to $N$ for $N$-state systems, and for this reason the HSEOM is computationally much more efficient than the conventional HEOM, in particular for larger systems such as a system described in a single \cite{SakuraiJPC11,KatoJPCB13,SakuraiJPSJ13,IkedaJCP17} and multidimensional coordinate space\cite{IkedaCP18}.
In addition, the HSEOM approach enables us to simulate the zero-temperature case or sub-Ohmic spectral density case.}
At this stage, however, it is necessary to employ a large set of Bessel functions, in particular, to study long-time behavior, due to the time profiles of Bessel functions, with which longer time behavior can be described only by a longer series of functions. To fully take advantage of the scalability of HSEOM in comparison to the HEOM, an efficient truncation scheme and the introduction of more appropriate special functions to describe the longer time behavior are necessary. We leave these tasks to future studies.

\acknowledgments

The authors would like to thank A. Kato (Institute for Molecular Science) for a valuable comment regerding using the spectral density with the circular cutoff. The financial support from a Grant-in-Aid for Scientific Research (A26248005) from the Japan Society for the Promotion of Science is acknowledged.

\appendix
\makeatletter
  \renewcommand{\theequation}{\Alph{section}.\arabic{equation}}{\@addtoreset{equation}{section}}
\makeatother
\makeatletter
  \renewcommand{\thefigure}{\Alph{section}.\arabic{figure}}{\@addtoreset{figure}{section}}
\makeatother
\makeatletter
  \renewcommand{\thetable}{\Alph{section}.\arabic{table}}{\@addtoreset{table}{section}}
\makeatother
\section{Path integral representation for a half-spin system}
\label{sec:spinPI}
We consider a half-spin system described by the operators $\hat{H}_{S}, \hat{V}, \hat{A}$, and $\hat{B}$ as the functions of Pauli matrices. The two-body correlation function $\Psi_{AB}(t;t')$ is evaluated in a path-integral form with the aid of a coherent spin state $\ket{\mu}$, given by\cite{SpinCoherent}
\begin{align}
\ket{\mu} = \frac{1}{\sqrt{1+|\mu|^2}}(\ket{1} + \mu \ket{0}),
\end{align}
where  $\mu$ is a complex number, and $\hat{\sigma}^{z} = \ket{1}\bra{1} - \ket{0}\bra{0}, \hat{\sigma}^{x} = \ket{1}\bra{0} + \ket{0}\bra{1}$, and $\hat{\sigma}^{y} = -i (\ket{1} \bra{0} - \ket{0}\bra{1})$ are the Pauli matrices.
The inner product and the completeness relation are expressed as
\begin{align}
\braket{\lambda|\mu} = \frac{1+\lambda^{*} \mu}{\sqrt{(1+|\lambda|^2)(1+|\mu|^2)}}, \\
\int d^{2} \mu \frac{2}{\pi (1+|\mu|^2)^2} \ket{\mu} \bra{\mu} = \hat{1}.
\end{align}

Then the matrix elements in terms of the coherent states are evaluated as
\begin{gather*}
\braket{\lambda|\hat{\sigma}^{z}|\mu} = \braket{\lambda|\mu} \frac{1-\lambda^{*}\mu}{1+\lambda^{*}\mu}, \\
\braket{\lambda|\hat{\sigma}^{x}|\mu} = \braket{\lambda|\mu} \frac{\lambda^{*} + \mu}{1+\lambda^{*}\mu}, \:
\braket{\lambda|\hat{\sigma}^{y}|\mu} = i\braket{\lambda|\mu} \frac{\lambda^{*} - \mu}{1+\lambda^{*}\mu}.
\end{gather*}
The system Hamiltonian $\hat{H}_{S}$ is then expressed as
$\braket{\lambda|\hat{H}_S|\mu} = \braket{\lambda|\mu} H_S(\lambda^{*}, \mu)$. The functional integrals and the initial state are expressed as
\begin{widetext}
\begin{gather*}
\int dq_{i} \int dq'_{i} \to \int d^{2} \mu_{i}\frac{2}{\pi(1+|\mu_{i}|^2)^2} \int d^{2} \mu'_{i} \frac{2}{\pi(1+|\mu'_{i}|^2)^2}, \\
\int \mathcal{D} [\tilde{q} (\cdot)] \to \int \mathcal{D}^2 [\tilde{\mu} (\cdot)]
 = \int d^{2} \mu \frac{2}{\pi(1+|\mu|^2)^2} \lim_{N \to \infty} \prod_{j=1}^{N-1} \left(\int d^{2} \mu_{j} \frac{2}{\pi(1+|\mu_{j}|^2)^2}
 \int d^{2} \mu'_{j} \frac{2}{\pi(1+|\mu'_{j}|^2)^2}\right), \\
\braket{q_{i}|\hat{\rho}_{S}(0)|q'_{i}} \to \braket{\mu_{i}|\hat{\rho}_{S}(0)|\mu'_{i}}.
\end{gather*}
The Lagrangian of the system is defined by the following equation:
\begin{align*}
\int_{C} d \tau L_{S} (\tilde{\mu}^{*}, \tilde{\mu}, \tau) & = \int_{0}^{t} d\tau L_{S}(\mu^{*}, \mu, \tau) - \int_{0}^{t} d\tau L^{\dagger}_{S}(\mu^{\prime*}, \mu', \tau) \\
& = \lim_{N \to \infty}\left\{ \Delta t \sum_{j=1}^{N} \left(-i \hbar \frac{\ln\braket{\mu_{j}|\mu_{j-1}}}{\Delta t} - H_{S}(\mu_{j}^{*}, \mu_{j-1})\right) - \Delta t \sum_{j=1}^{N} \left(i \hbar \frac{\ln\braket{\mu_{j-1}^{\prime *}|\mu'_{j}}}{\Delta t} - H_{S}(\mu_{j-1}^{\prime *}, \mu'_{j})\right) \right\},
\end{align*}
\end{widetext}
where $\Delta t = t/N, \mu_{0} = \mu_{i}, \mu'_{0} = \mu'_{i}$, and $\mu_{N} = \mu'_{N} = \mu$.

The coherent-state representations of $\hat{A}, \hat{B}$, and $\hat{V}$ are defined in the same way as the system Hamiltonian.

The HSEOM  for half-spin systems are derived in the same form as Eq. \eqref{HSEOMC1} with the aid of the discrete path-integral form by differentiating RWF and AWFs in the same way as in the HEOM of the boson coherent form\cite{TanimuraJPSJ06}.

\end{document}